\documentclass[aps, english,pra,superscriptaddress,showpacs,twocolumn,10pt]{revtex4}

\usepackage{amsfonts,amssymb,amsmath}
\usepackage{graphics,graphicx,epsfig}
\usepackage{amsthm}
\usepackage{epstopdf}

\def\identity{\leavevmode\hbox{\small1\kern-3.8pt\normalsize1}}

\newcommand{\ket}[1]{\left | #1 \right\rangle}
\newcommand{\bra}[1]{\left \langle #1 \right |}

\newcommand{\proj}[1]{\ket{#1}\bra{#1}}
\renewcommand{\epsilon}{\varepsilon}

\makeatletter

\@ifundefined{textcolor}{}
{%
 \definecolor{BLACK}{gray}{0}
 \definecolor{WHITE}{gray}{1}
 \definecolor{RED}{rgb}{1,0,0}
 \definecolor{GREEN}{rgb}{0,1,0}
 \definecolor{BLUE}{rgb}{0,0,1}
 \definecolor{CYAN}{cmyk}{1,0,0,0}
 \definecolor{MAGENTA}{cmyk}{0,1,0,0}
 \definecolor{YELLOW}{cmyk}{0,0,1,0}
 }

\makeatother

\usepackage{babel}

\begin{document}

\title{Experimental Schmidt Decomposition and State Independent Entanglement Detection}

\author{Wies\l aw~Laskowski}

\affiliation{Institute of Theoretical Physics and Astrophysics, University of Gda\'nsk, PL-80-952 Gda\'nsk, Poland}
\affiliation{Max-Planck-Institut f\"ur Quantenoptik, Hans-Kopfermann-Strasse 1, D-85748 Garching, Germany}
\affiliation{Department f\"ur Physik, Ludwig-Maximilians-Universit\"at, D-80797 M\"unchen, Germany}

\author{Daniel~Richart}
\author{Christian~Schwemmer}

\affiliation{Max-Planck-Institut f\"ur Quantenoptik, Hans-Kopfermann-Strasse 1, D-85748 Garching, Germany}
\affiliation{Department f\"ur Physik, Ludwig-Maximilians-Universit\"at, D-80797 M\"unchen, Germany}

\author{Tomasz~Paterek}

\affiliation{Centre for Quantum Technologies, National University of Singapore, 3 Science Drive 2, 117543 Singapore, Singapore}

\author{Harald~Weinfurter}

\affiliation{Max-Planck-Institut f\"ur Quantenoptik, Hans-Kopfermann-Strasse 1, D-85748 Garching, Germany}
\affiliation{Department f\"ur Physik, Ludwig-Maximilians-Universit\"at, D-80797 M\"unchen, Germany}

\begin{abstract}
We introduce an experimental procedure for the detection of quantum entanglement of an unknown quantum state with a small number of measurements. The method requires neither a priori knowledge of the state nor a shared reference frame between the observers and can thus be regarded as a perfectly state independent entanglement witness. 
The scheme starts with local measurements, possibly supplemented with suitable filtering, which essentially establishes the Schmidt decomposition for pure states. Alternatively we develop a decision tree which reveals entanglement within few steps.
These methods are illustrated and verified experimentally for various entangled states of two and three qubits.
\end{abstract}


\maketitle

\emph{Introduction.}---Entanglement is the distinguishing feature of quantum mechanics and it is the most important resource for quantum information processing \cite{NIELSEN, HORODECCY}. For any experiment it is thus of utmost importance to easily reveal entanglement, best with as little effort as possible. Common methods suffer from disadvantages. On the one hand, employing the Peres-Horodecki criterion \cite{PERES,HoroNPT} or evaluating entanglement measures, one can identify entanglement in arbitrary states, however, it requires full state tomography. On the other hand, various entanglement witnesses \cite{HoroNPT,BOURENNANE_ETAL_2004,TERHAL2000,LEWENSTEIN_ETAT_2000,BRUSS_ETAL_2002,GH2003,GT2009} can be determined with much fewer measurements but give conclusive answers only if the state under investigation is close to the witness-state, i.e., they require a priori knowledge.

\vspace{-0.1cm}Recently, it has been shown that the existence of entanglement can be inferred from analyzing correlations between the measurement results on the subsystems of a quantum state. Only if the state is entangled, the properly weighted sum of correlations will overcome characteristic thresholds \cite{GEOM_SEP}.
Here we further develop this approach to obtain a simple and practical method to detect entanglement of all pure states and some mixed states by measuring only a small number of correlations.
Since the method is adaptive it does not require a priori knowledge of the state nor a shared reference frame between the possibly remote observers and thus greatly simplifies the practical application.
We describe two schemes where the first one essentially can be seen as a direct implementation of Schmidt decomposition, which identifies the maximal correlation directly. For bipartite pure  systems, this approach can conceptually be divided into two stages:
(i) calibration that establishes the experimental Schmidt decomposition \cite{SCHMIDT,PERES_BOOK} of a pure state by local measurements and suitable filtering  and (ii) two correlation measurements to verify the entanglement criterion. The second scheme shows how to use a decision tree to obtain a rapid violation of the threshold thereby identifying entanglement.

\emph{Entanglement criterion.}---For a two-qubit quantum state $\rho$, 
Alice and Bob observe correlations between their local Pauli measurements $\sigma_{k}$ and $\sigma_l$, respectively. They are defined as the expectation values of the product of the two measurements,
$T_{kl} = \mbox{Tr}[\rho (\sigma_{k} \otimes \sigma_{l})]$, with the so called correlation tensor elements $T_{kl} \in [-1,1]$. 
The local values $T_{k0}$ ($T_{0l}$), with $\sigma_{0}$ being the identity operator, form the local Bloch vector of Alice (Bob).
Using these measurements a sufficient condition for entanglement can be formulated as \cite{GEOM_SEP, HORODECKI1996}:
\begin{equation}
\sum_{k,l = x,y,z} T_{kl}^2 > 1 \quad \Rightarrow \quad \rho \textrm{ is entangled}.
\label{SIMPLE_CRIT}
\end{equation}
For pure states this is also a necessary condition, while for mixed states care has to be taken. For those the likelihood for detecting the entanglement 
decreases with purity. An extension of (\ref{SIMPLE_CRIT}) can generally identify entanglement of an arbitrary mixed state, however, then loosing the state independence \cite{GEOM_SEP}.
Note two important facts. First, Eq.~\eqref{SIMPLE_CRIT} can be seen as a state independent entanglement witness derived without any specific family of entangled states in mind. 
Second, to test whether the state is entangled, it is sufficient to break the threshold, i.e., it is neither required to measure all correlations nor to compute the density matrix \footnote{Although the effort to evaluate (\ref{SIMPLE_CRIT}) generally scales as the one for full state tomography, we can directly evaluate the criterion from raw data without any further numerics required to reconstruct a physical density matrix \cite{MAX_LI}}. Rather, it is now the goal to find strategies which minimize the number of correlation measurements. We show how this can be done by a particularly designed decision tree, or by identifying a Schmidt decomposition from local results and filtering when necessary.

\emph{Schmidt decomposition.}---Consider pure two qubit states.
Any such state has a Schmidt decomposition
\begin{equation}
\ket{\psi_S} = \cos \theta \ket{a} \ket{b} + \sin \theta \ket{a_\perp} \ket{b_\perp}, \quad \theta \in [0,\tfrac{\pi}{4}],
\label{SCHMIDT}
\end{equation}
where the coefficients are real and the local bases $\{\ket{a},\ket{a_\perp}\}$ and $\{\ket{b}, \ket{b_\perp}\}$ are called the Schmidt bases.
Once the bases are known, Alice constructs her local measurements
$\sigma_{z'} = \ket{a} \bra{a} - \ket{a_\perp} \bra{a_\perp}$ and $\sigma_{y'} = i \ket{a_\perp} \bra{a} -i \ket{a} \bra{a_\perp}$,
and so does Bob in analogy.
They can now detect entanglement with only {\emph two} correlation measurements because $T_{z^{'}z^{'}}^2 + T_{y^{'}y^{'}}^2 = 1 + \sin^2 2 \theta > 1$ for all pure entangled states.
Note, the laboratories are not required to share a common reference frame.

In order to extract the Schmidt bases from experimental
data one starts with local measurements, determining the local Bloch vectors $\vec \alpha ~(\vec \beta)$ of Alice (Bob) (those elements are related to the correlation tensor coefficients via $\alpha_i = T_{i0}/\sqrt{T_{x0}^2+T_{y0}^2+T_{z0}^2}$).
We consider two cases. First, suppose that a pure state has non-vanishing local Bloch vectors.
Their directions define the Schmidt bases of Alice and Bob up to a global phase $\phi$.
Writing these bases in the computational basis
\begin{eqnarray}
\ket{a} & = & \cos \xi_{A} \ket{0} + e^{i \varphi_{A}} \sin \xi_{A} \ket{1}, \label{schmidtAlice} \nonumber \\
\ket{a_\perp} & = & \sin \xi_{A} \ket{0} - e^{i \varphi_{A}} \cos \xi_{A} \ket{1}, \label{schmidtAliceP} \nonumber\\
\ket{b} & = & \cos \xi_{B} \ket{0} + e^{i \varphi_{B}} \sin \xi_{B} \ket{1}, \label{schmidtBob} \nonumber \\
\ket{b_\perp} & = & e^{i \phi}(\sin \xi_{B} \ket{0} - e^{i \varphi_{B}} \cos \xi_{B} \ket{1}), \label{schmidtBobP}
\end{eqnarray}
we see that the required coefficients can be inferred directly from the local Bloch vectors, 
$\vec \alpha  = (\sin 2 \xi_{A} \cos \varphi_{A}, \sin 2 \xi_{A} \sin \varphi_{A}, \cos 2\xi_{A})$
on Alice's side, and similarly for Bob. 
The global phase of $\ket{b_\perp}$ shows up as the relative phase in the decomposition (\ref{SCHMIDT}), i.e. $\ket{\psi_S} = \cos \theta \ket{a} \ket{b} + \sin \theta e^{i \phi} \ket{a_\perp} |\tilde b_\perp \rangle$ (with $\ket{b_\perp} = e^{i \phi} \ket{\tilde b_\perp}$). It can be determined, for example, from the $T_{yy}$ correlation as $\cos \phi = T_{yy}/\sqrt{1 - T_{x0}^2 - T_{y0}^2 - T_{z0}^2}$. 
If Bob would use the basis $\{\ket{b}, |\tilde b_\perp \rangle \}$ to build his observables $\sigma_{z''}$ and $\sigma_{y''}$, the corresponding correlations $T_{y'y''} = \sin 2\theta \cos\phi$ would vanish for $\cos\phi=0$ and the two measurements $T_{z'z''}$ and $T_{y'y''}$ would not suffice to detect entanglement. In such a case, however, the other two correlations, $T_{x'y''}$ and $T_{y'x''}$, are non-zero, and can be used to reveal entanglement. Therefore, the determination of $\phi$ in the calibration is not essential if one accepts possibly one more correlation measurement.

Second, in case of vanishing local Bloch vectors, the pure state under consideration $\ket{\psi_m}$ is maximally entangled  and admits infinitely many Schmidt decompositions. In order to truly prove entanglement Bob can thus freely choose some basis, say computational basis, for which the state will now be of the form $\ket{\psi_m}  = \frac{1}{\sqrt{2}}(\ket{a} \ket{0} + \ket{a_\perp} \ket{1})$. 
The basis of Alice can be found after \emph{filtering} by Bob in his Schmidt basis: $F = \ket{0} \bra{0} + \epsilon \ket{1} \bra{1}$ (for an actual implementation see experimental section).
When Bob informs Alice that his detector behind the filter clicked, the initial state becomes
\begin{equation}
(\openone \otimes F) \ket{\psi_m} \to \frac{1}{\sqrt{1 + \epsilon^2}} (\ket{a} \ket{0} + \epsilon \ket{a_\perp} \ket{1}).
\label{filtering}
\end{equation}
Note that due to filtering a nonvanishing local Bloch vector emerges for Alice. Thus the respective Schmidt basis can be found with the method described above and be used for the evaluation of $T^2_{z'z}+T^2_{y'y}$.

\begin{figure}
\includegraphics[width=0.28\textwidth]{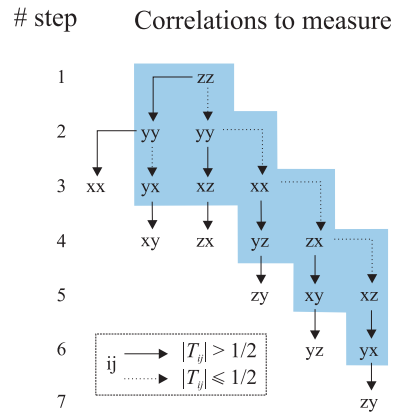}
\caption{The decision tree for efficient two-qubit entanglement detection. No shared reference frame is required between Alice and Bob, i.e. they choose their local $x,y,z$ directions randomly and independently, which effectively gives rise to a basis $\{x_A, y_A, z_A$\} for Alice and $\{x_B, y_B, z_B\}$ for Bob (not detailed in the Figure or the main text).
The scheme starts with measuring $T_{zz}$ and follows at each step along the dashed arrow if the modulus of correlation is less than $\frac{1}{2}$ and along the continuous arrow otherwise. The algorithm succeeds as soon as  $\sum T_{ij}^2 >1$.  The measurements in blue shaded area suffice to detect all maximally entangled pure states with Schmidt bases vectors along $x$,$y$ or $z$.}
\label{FIG_DT}
\end{figure}

\emph{Decision tree.}---Our second algorithm for entanglement detection does not even require any calibration and also applies directly to mixed states.
Alice and Bob choose three orthogonal local directions $x,y$ and $z$ independently from each other and agree to only measure correlations along these directions.
In Fig. \ref{FIG_DT} we show exemplarily which correlations should be measured in order to detect entanglement in a small number of steps.
Starting with a measurement of $T_{zz}$, one continues along the solid (dotted) arrow, if the correlation is higher (lower) than some threshold value (e.g. 1/2 
in Fig.~\ref{FIG_DT}). 
The tree is based on the principle of correlation complementarity \cite{CORRELATION_COMPLEMENTARITY,TG2005, WW_CC_2008, WW_CC_2010}:
in quantum mechanics there exist trade-offs for the knowledge of dichotomic observables with corresponding anti-commuting operators.
For this reason, if the correlation $|T_{zz}|$ is big, correlations $|T_{zx}|,|T_{zy}|,|T_{xz}|$ and $|T_{yz}|$ have to be small 
because their corresponding operators anti-commute with the operator $\sigma_{z} \otimes \sigma_{z}$.
Therefore, the next significant correlations have to lie in the $xy$ plane of the correlation tensor and thus the tree continues with a measurement of the $T_{yy}$ correlation.
This concept can be generalized to multiqubit states, a decision tree for three qubits is given in the Appendix.
The number of detected states grows with the number of steps through the decision tree. Since condition (1) is similar to the purity of a state, obviously, the scheme succeeds the faster the more pure a state is (see Appendix for detailed analysis). Varying the threshold value does not lead to any significant changes in the statistic of detected states.

Finally, we connect both methods discussed here for the analysis of multiqubit states. A numerical simulation for pure states reveals that the correlation measurement along local Bloch vectors 
gives correlations close to the maximal correlations in more than $80\%$ of the cases.
Therefore, these local directions give an excellent starting point for the decision tree.

\begin{figure}[!t]
\includegraphics[width=0.4\textwidth]{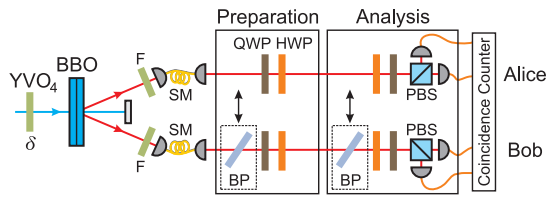}
\caption{Scheme of the experimental setup.
The state $\ket{\Psi}=\frac{1}{\sqrt{2}}(\ket{H}\ket{H}+e^{i \delta}\ket{V}\ket{V})$ is created by type I SPDC process. An Yttrium Vanadate crystal (YVO$_{4}$) is used to manipulate the phase $\delta$ of the prepared state. For preparation and analysis of the state half- (HWP) and quarter (QWP) waveplates are employed. Brewster plates (BP) can be introduced to make the state asymmetric and to perform the filter operation, respectively.}
\label{source}
\end{figure}

\emph{Experiment.}--- For the demonstration of these new simple analysis methods we first use two photon polarization entangled states. In the following we will thus replace the computational basis states by horizontal ($\ket{0} \rightarrow \ket{H}$) and vertical ($\ket{1} \rightarrow \ket{V}$) linear polarization, respectively. The photon source (see Fig. \ref{source}) is based on the process of spontaneous parametric down conversion (SPDC) using a pair of crossed type I cut $\beta$-Barium-Borate (BBO) crystals pumped by a CW laser diode at a wavelength of $\lambda_{pump}=402$nm with linear polarization of $45^{\circ}$. It emits pairs of horizontally and vertically polarized photons which superpose to the state $\ket{\Psi}=\frac{1}{\sqrt{2}}(\ket{H}\ket{H}+e^{i\delta}\ket{V}\ket{V})$ \cite{KWIAT}.
The spectral bandwidth of the photons is reduced to $5$nm using interference filters and two spatial emission modes are selected by coupling the photon pairs into two separate single mode fibers. 

For the purpose of preparing any pure two-qubit state, the polarization of each photon can be rotated individually by a set of quarter- (QWP) and half (HWP) waveplates in each mode. By tilting an Yttrium Vanadate crystal (YVO$_{4}$) in front of the BBOs, the relative phase $\delta$ between the photon pairs can be set. Additionally, the state can be made asymmetric by removing a portion of vertically polarized light in one spatial mode with a Brewster plate (BP).
In the last step of the experiment, the polarization of each photon is analyzed with additional half- and quarter waveplates and projection on $\ket{H}$ and $\ket{V}$ using a polarizing beam splitter (PBS). The local filtering of a maximally entangled state can be accomplished by placing a Brewster plate in front of the analysis waveplates.
This Brewster plate reflects with a certain probability vertically polarized photons and together with detection of a photon behind the Brewster plate implements the filtering operation (\ref{filtering}). Finally, the photons are detected by fiber-coupled single photon detectors connected to a coincidence logic.

\begin{figure}[!t]
\includegraphics[width=0.4\textwidth]{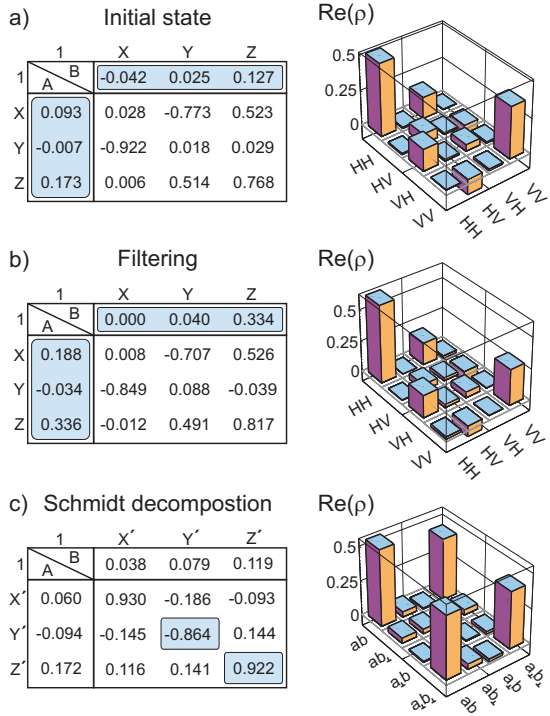}
\caption{Demonstration of Schmidt decomposition of an maximally entangled state prepared in unknown bases. The correlation tensor and corresponding density matrix are depicted for the unknown state a), after applying local filtering b) and for the state analyzed in the Schmidt bases c). It is important to note that only the blue shaded elements of the correlation tensors  will be measured as this suffices to prove entanglement. The full correlations and the density matrices of the corresponding states are only shown for completeness and didactical reasons.}
\label{schmidtsym}
\end{figure}

\emph{Experimental Schmidt decomposition.}--- Let us consider the state shown in figure \ref{schmidtsym}a).
The protocol starts with Alice and Bob locally measuring the polarization of the photons enabling them to individually determine the local Bloch vectors. For high efficiencies, e.g., possible in experiments with atoms or ions, the local measurements can indeed be done independently 
\footnote{Due to the low efficiency of the SPDC setup we have to use coincidence counts here.
Using only the single counts, the local Bloch vector of Alice is slightly different. In this example we obtain $T_{0l} = (-0.073, -0.091, -0.059)$ and $T_{k0} =  (0.031,0.041, 0.037)$.}.
If nonvanishing local Bloch vectors can be identified, one can proceed to the next step. For the example here, the local expectation values are close to zero and filtering has to be applied. 
By using a Brewster plate in front of Bob's analysis waveplate and under the condition that the filtering operation is successful, local Bloch vectors emerge (see Fig. \ref{schmidtsym}b) \footnote{This in consequence means, that for the filtering Alice and Bob have to communicate with each other and hence this measurement is \textit{not} local.}, in this case we obtain $T_{0l} = (0.000,0.040,0.334)$ and $T_{k0} = (0.188,-0.034,0.336)$. 

In the next step Alice and Bob use their local Bloch vectors to realign their analyzers to the new local Schmidt bases $\{ \ket{a}$, $\ket{a_{\perp}} \}$ and $\{ \ket{b}$, $| \tilde b_{\perp}\rangle \}$, respectively.
This diagonalizes the correlation tensor as depicted in figure \ref{schmidtsym}c). Therefore, it is only necessary to measure $T_{z'z''}=0.922 \pm0.015$ and $T_{y'y''}=-0.864\pm0.015$ to prove entanglement since $T_{z'z''}^2 + T_{y'y''}^2 = 1.597 \pm 0.038 > 1$. Hence, $2 \times 3$ \textit{local} measurements are needed in the first step of the algorithm, if necessary three combined measurements are needed for filtering, and finally only two correlation measurements have to be performed for entanglement detection.

\begin{figure}[ht]
\includegraphics[width=0.4\textwidth]{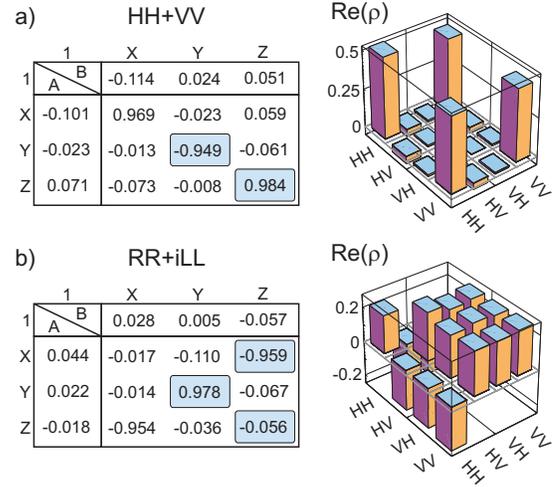}
\caption{Correlation tensors and density matrices of the experimental realization of two different states. The imaginary parts of the density matrices are negligible and therefore skipped. Using the decision tree, only the blue shaded correlations have to be measured for detecting entanglement. The errors of the correlations are $< 0.015$ for a) and $<0.023$ for b).}
\label{HHVVandRRLL}
\end{figure}

\emph{Application of the decision tree.}--- In order to demonstrate the application of the decision tree we will apply it to three states. For the first state $\frac{1}{\sqrt{2}}(\ket{H}\ket{H}+\ket{V}\ket{V})$, whose correlation tensor is depicted in \ref{HHVVandRRLL}a) the decision tree (see Fig. \ref{FIG_DT}) starts with the measurement of the correlation $T_{zz}=0.980\pm0.015$ and continues with $T_{yy}=-0.949\pm0.015$. These two measurements already prove entanglement since $T_{zz}^2 + T_{yy}^2 = 1.869\pm0.041 > 1$. 
For a second state, $\frac{1}{\sqrt{2}}(\ket{R}\ket{R}+i\ket{L}\ket{L})$, (see Fig. \ref{HHVVandRRLL}b),  we obtain a correlation of $T_{zz}=-0.056\pm0.015$, close to zero. Consequently, the next steps according to our algorithm (Fig. \ref{FIG_DT}) are to determine the correlation $T_{yy}=0.978\pm0.015$ followed by $T_{xz}=-0.959\pm0.015$ , with their squares adding up to a value of $1.879\pm0.041 > 1$ and hence proving entanglement. As a last example we consider the initial state of Fig. \ref{schmidtsym}. According to our decision tree we need to measure $T_{zz} = 0.768 \pm 0.015$, $T_{yy} = 0.018 \pm 0.015$ and $T_{yx} = -0.922 \pm 0.015$, thus giving a value of $1.440 \pm 0.036 >1$ and proving entanglement with only three steps.

\begin{figure}[!t]
\includegraphics[width=0.4\textwidth]{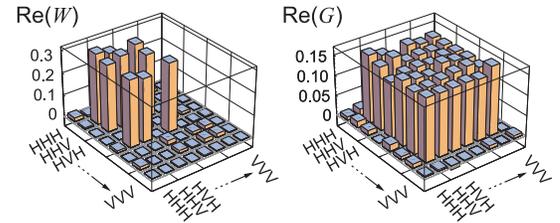}
\caption{Density matrices of the experimental realization of the $G$ and $W$ state. The corresponding fidelities are equal to 92.23\% and 89.84\%.}
\label{multi}
\end{figure}

\emph{Many qubits.}---For the demonstration of multiqubit entanglement detection, we use two three-photon polarization entangled states: the $W$ state \cite{W} and the $G$ state \cite{G} (see Fig. \ref{multi}).
In order to observe these states, a collinear type II SPDC source together with a linear setup to prepare the four-photon Dicke state $D_4^{(2)}$ is used \cite{KRISCHEK, KIESEL}.
Once the first photon is measured to be vertically polarized, the other three photons are projected into the $W$ state.
Similarly, the three-photon $G$ state is obtained if the first photon is measured to be $+45^\circ$ polarized.

The protocol for entanglement detection starts with observers locally measuring the polarization of the photons enabling them to individually determine the local Bloch vectors. 
For the $G$ state we obtain 
$T_{i00}=(0.636,-0.008, -0.015)$, 
$T_{0j0}=(0.623,-0.092,0.010)$ and
$T_{00k}=(0.636, 0.070,0.022)$. 
The local Bloch vectors suggest that the correlation $T_{xxx}$ is big. Therefore the decision tree starts with the measurement of $T_{xxx} =0.904 \pm 0.025$ 
and continues with $T_{xzz} = -0.578 \pm 0.025$ (see Fig. 7 in the Appendix). These two measurements already prove entanglement because $T_{xxx}^2 + T_{xzz}^2 = 1.152 \pm 0.038 > 1$.
For the $W$ state, the local Bloch vectors: 
$T_{i00}=(0.016, -0.070, 0.318)$, 
$T_{0j0}=(-0.010, -0.073, 0.308)$ and
$T_{00k}=(-0.011, -0.0547, 0.319)$ suggest that now the correlation $T_{zzz}$ is big. Indeed, we observe $T_{zzz} = -0.882 \pm 0.025$.
The decision tree is the same as above but with local axes renamed as follows $x \to z \to y \to x$.
Therefore, the second measurement has to be $T_{zyy}$. With $T_{zyy} = 0.571 \pm 0.025 $ we again prove entanglement as $T_{xxx}^2 + T_{zyy}^2 = 1.104 \pm 0.037 > 1$.

\emph{Conclusions.}---
We discussed and experimentally implemented two methods for fast entanglement detection for states about which we have no a priori knowledge. They are well suited for quantum communication schemes as the parties do not have to share a common reference frame, making the scheme insensitive to a rotation of the qubits during their transmission to the distant laboratories. The two methods use a particularly simple and practical entanglement identifier \cite{GEOM_SEP}.
One of them can be seen as experimental Schmidt decomposition and the other establishes a sequence of correlation measurements leading to entanglement detection in a small number of steps. 

We thank M. \.Zukowski for stimulating discussions.
This work is supported by the EU project QESSENCE, the DAAD/MNiSW, the DFG-Cluster of Excellence MAP,
and by the National Research Foundation and Ministry of Education in Singapore.
WL is supported by the MNiSW Grant no. N202 208538 and by
the Foundation for Polish Science. CS thanks QCCC of the Elite Network of Bavaria for support.

\appendix*

\section{Decision Tree}

This section is devoted to study the efficiency of the decision tree algorithm described in the main text.
Some results in this section are analytical and some are numerical.
In all our numerical investigations (unless explicitly stated otherwise) we used the decision tree of the main text (for two qubits)
and in cases when going through the whole tree did not reveal entanglement 
we augmented it with additional measurements of those correlations which were not performed until that moment.
The order of the additional measurements also results from the correlation complementarity (anti-commutation relations)~\cite{CORRELATION_COMPLEMENTARITY}.
With every remaining measurement we associate the ``priority'' parameter
\begin{equation}
P_{ij} = \sum_{k \ne i} P_{ij}(T_{kj}) + \sum_{l \ne j} P_{ij}(T_{il}),
\end{equation}
that depends on the measured correlation tensor elements of the decision tree in the following way
\begin{equation}
P_{ij}(T_{mn}) = \Big\{
\begin{array}{rl}
T_{mn}^2 & \textrm{  if } T_{mn} \textrm{ was performed before,} \\
0 & \textrm{  else.}
\end{array}
\end{equation}
According to the correlation complementarity there is a bigger chance that this correlation is significant if the value of the corresponding parameter is small.
Therefore, the correlations $T_{ij}$ with lower values of $P_{ij}$ are measured first. 

Let us consider the following example. 
The measured correlations of the decision tree are as follows: $T_{zz}=0.7$, $T_{xx}=0.1$, and $T_{yy}=0.4$. 
Therefore, $P_{xy} = P_{yx} = T_{xx}^2 + T_{yy}^2 = 0.17$ , $P_{xz} = P_{zx} = T_{xx}^2 + T_{zz}^2 = 0.5$, and $P_{zy} = P_{zy} = T_{zz}^2 + T_{yy}^2 = 0.65$.
Accordingly the order of the remaining measurements is as follows: 
first measure $xy$, then $yx$, next $xz,zx,yz$ and $zy$.

\subsection{Two qubits}

\subsubsection{Werner states}

As an illustration of how the decision tree works for a well-known class of mixed states we first consider Werner states.
It turns out that not all entangled states of the family can be detected.

Consider a family of states
\begin{equation}
\rho = p \proj{\psi^-} + (1-p) \frac{1}{4} \openone,
\label{WERNER_STATE}
\end{equation}
where $\ket{\psi^-} = \frac{1}{\sqrt{4}}(\ket{01} - \ket{10})$ is the Bell singlet state,
$\frac{1}{4} \openone$ describes the completely mixed state of white noise, and $p$ is a probability.
Its correlation tensor, written in the same coordinate system for Alice and Bob, is diagonal with entries $T_{xx} = T_{yy} = T_{zz} = -p$, arising from the contribution of the entangled state.
The states (\ref{WERNER_STATE}) are entangled if and only if $p > \frac{1}{3}$,
whereas the decision tree reveals that these states are entangled for $p > \frac{1}{\sqrt{3}} \approx 0.577$.
This is because only three elements contribute to the criterion.
Note that the value of the decision parameter is irrelevant here.
Furthermore, a random choice of local coordinate directions does not help.
Although more steps would be involved in the decision tree the sum of squared correlations is invariant under local unitary operations 
and therefore only for $p > \frac{1}{\sqrt{3}}$ entanglement is detected for the Werner state.

\subsubsection{Entangled state mixed with colored noise}

An exemplary class of density operators for which the decision tree detects all entangled states is provided by:
\begin{equation}
\gamma = p \proj{\psi^-} + (1-p) \proj{01},
\end{equation}
where entanglement is mixed with colored noise $\ket{01}$ bringing anti-correlations along local $z$ axes.
This state has the following non-vanishing elements of its correlation tensor $T_{xx} = T_{yy} = -p$ and $T_{zz} = -1$.
Therefore, the decision tree allows detection of entanglement for this class of states in two steps.
Note that the state is entangled already for an infinitesimal admixture of the Bell singlet state as can be shown for a random choice of local coordinate systems.

\subsubsection{Random mixed states}

\begin{figure}[h]
\includegraphics[width=0.40\textwidth]{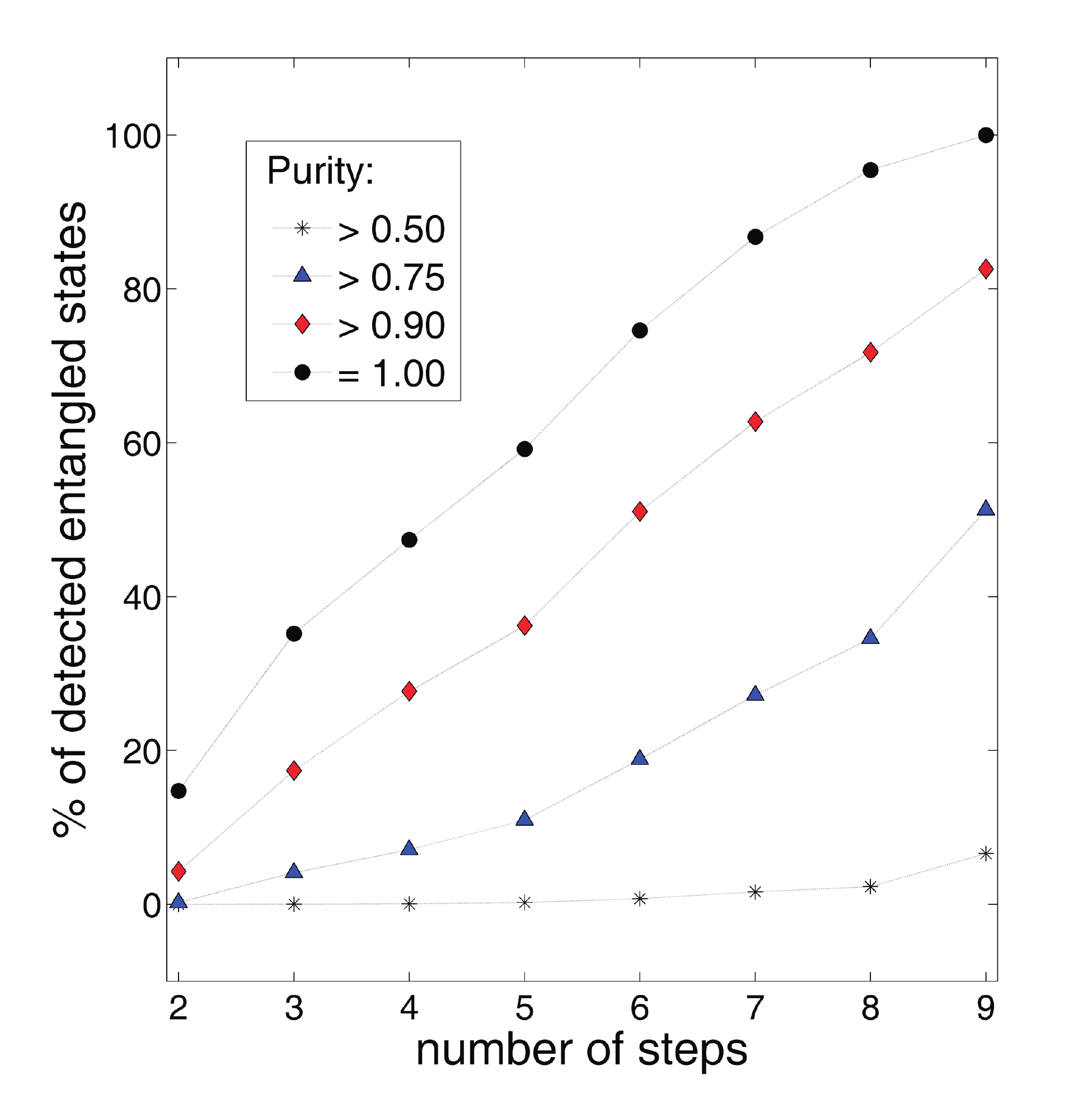}
\caption{Efficiency of the decision tree for two qubit random mixed states.}
\label{SI_FIG_RANDOM_MIXED}
\end{figure}

Fig. \ref{SI_FIG_RANDOM_MIXED} shows how the efficiency of the algorithm grows with the purity of tested states.
The efficiency is measured by the fraction of detected entangled states obtained from extensive Monte Carlo sampling in the two qubit state space. 
For nine steps the algorithm detects all the pure states. This is expected because Eq. (1) of the main text is a necessary and sufficient condition 
for the detection of entanglement of pure states.

\subsection{Many qubits}

\subsubsection{Algorithm for generation of the tree}

The principle behind the decision tree is the correlation complementarity~\cite{CC}.
Correlation complementarity states that for a set of dichotomic anti-commuting operators $\{\alpha_1,\dots, \alpha_k \}$
the following trade-off relation is satisfied by all physical states:
\begin{equation}
T_{\alpha_1}^2 + \dots T_{\alpha_k}^2 \le 1,
\end{equation}
where $T_{\alpha_1}$ is the average value of observable $\alpha_1$ and so on.
Therefore, if one of the average values is maximal, $\pm 1$, the other anti-commuting observables have vanishing averages.
This motivates taking only sets of commuting operators as different branches of the decision tree.

\subsubsection{Decision tree for three qubits}

\begin{figure}[h]
\includegraphics[width=0.42\textwidth]{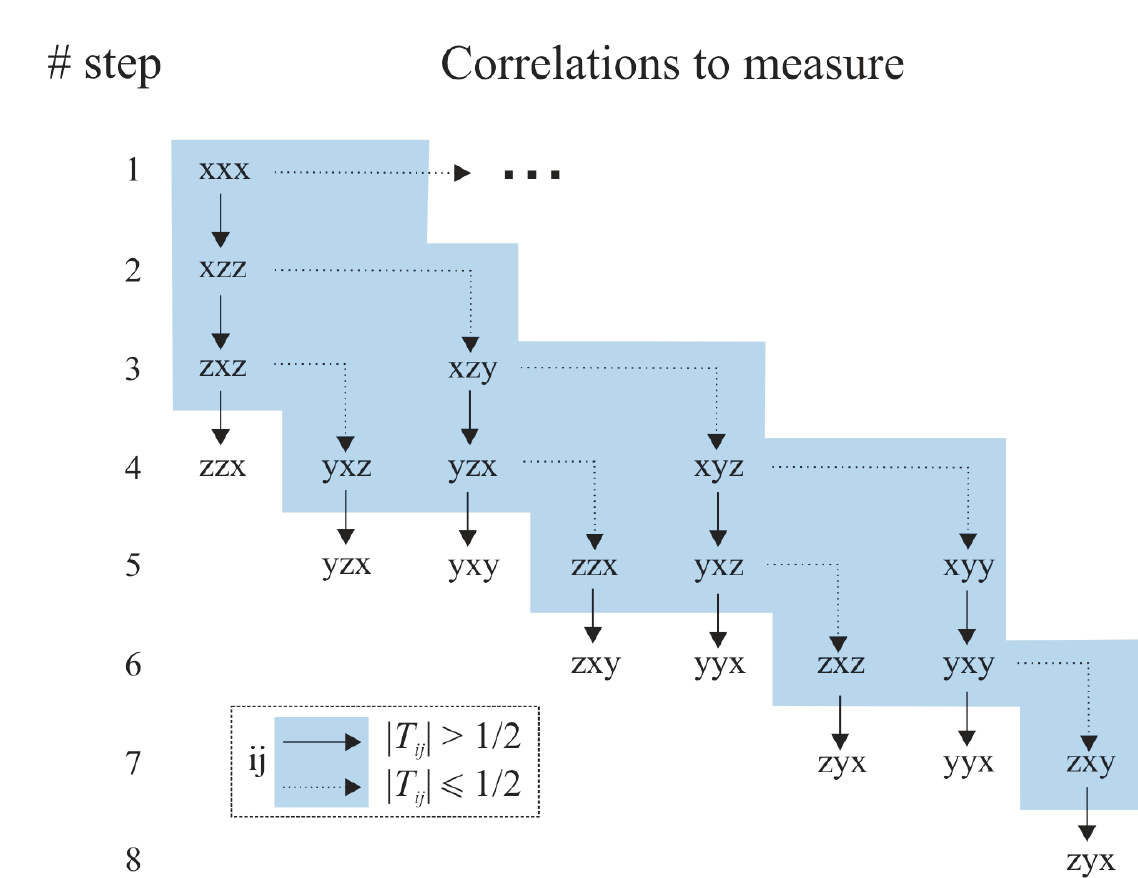}
\caption{One branch of the decision tree for three qubits.}
\label{DT_3QUBITS}
\end{figure}

This tree is an example of the application of the algorithm presented in the previous section.
Fig. \ref{DT_3QUBITS} shows only one branch of the whole tree.

A numerical simulation reveals that the correlation measurement along local Bloch vectors 
gives correlations close to the maximal correlations of a pure multi-qubit state in more than $80\%$ of the cases.
Therefore, these local directions give an excellent starting point of the decision tree.

Exemplarily, the branch begins with $T_{xxx}$ assuming that this correlation is big.
If the local Bloch vectors indicate correlation along a different directions 
it is advantageous to correspondingly change the elements of the decision tree (see main text).

\end{document}